# Some Long-Standing Quality Practices in Software Development

Sibonile Moyo
National University of Science and Technology
Bulawayo
Zimbabwe

email: sibonile.moyo@nust.ac.zw

**Abstract**
The desire to build quality software systems has been the focus of most software developers and researchers for decades. This has culminated in the design of practices that promote quality in the designed software. Originating from the inception of the traditional software development life cycle (SDLC), through to the object-oriented methods, Iterative development, and now the agile methods, these practices have persisted through different periods. Such practices play the same quality role regardless of the perspective of the software development process they are part of. In this paper we review three software development methods representative of the software development history, with the aim of i) identifying key quality practices, ii) identifying the quality role played by the practice in the method, and iii) noting those quality practices that have persisted through the software development history. The identified quality practices that have persisted throughout the history of the software development processes include prototyping, iterative development, incremental development, risk-driven development, phase planning, and phase retrospection. These results would be useful to method engineers who seek to design high-quality software development methods as these practices serve as candidates for inclusion in their development processes. Software development practitioners seeking to design quality software would also benefit from adopting these practices in developing their software.

1. Introduction

Software quality refers to the capability of a defined process to deliver a software artefact that meets its intended use, at the same time adhering to the generic 'ilities' of software (Moyo, 2020). This definition adopts a software process perspective of software quality (García-Mireles et al., 2015). The assumption in this definition is that software processes are designed with practices that build quality into the software. This has been the focus of most software development processes, that are built to navigate the generic software development life cycle systematically.

Having taught software engineering courses for more than two decades, the author has observed the emphasis in software development shift from a contractual, heavyweight software process to that of a flexible agile approach to building software. Regardless of whether the method under consideration is traditional, agile, or any hybrid of these, the goal in software development is to create an acceptable artifact that meets user requirements, and can easily be integrated with existing systems if need be (Karhapää et al., 2021). From an educational perspective, the aim is to equip the graduates of software engineering with the key skills required for the practice (Mbambo-Thatha & Moyo, 2014), taking into cognizance that such a graduate may find employment in any of the active software engineering fields of the day.

Over the years it has been interesting to note some practices and principles appearing in the several new methods that method engineers develop to improve software quality. While in some instances practices appear as they are, in some cases they appear with new names but refer to the same principle. In such cases what becomes pertinent is the role played by the practice in enhancing software quality (Kassab, 2015; Maxim & Kessentini, 2016). Regardless of the naming adopted, and the perspective taken, there are some software quality practices that have persisted over the years and continue to influence the quality of the designed software positively.

This paper gives an overview of those practices, techniques, and principles that have penetrated the various periods of the software engineering processes. We note the practice and highlight the methods that have incorporated it, and the impact of the practice on software quality. In Section 2, we present the views of other authors on the history of software quality. Section 3 outlines the method used in identifying the practices. Section 4 presents the results and Section 5 briefly discusses the findings and Section 6 concludes the paper.

2. Background and Related work

In seeking to understand the past in comparison with the present, researchers have discussed the history of software development in its various forms. Boehm (2006) presented key improvements in software processes between 1950 and 2010 and beyond. The author highlights milestones in the software development environment during the period, making future projections on what could still be applicable in the future. In this discussion the author highlights key practices that are worth carrying over into the future in a bid to enhance the quality of the developed software.

Boehm's efforts in ensuring that the successes of the past are not forgotten have been complemented by several researchers through systematic reviews in software engineering. Magdaleno et al. (2012) explored the differences among plan-driven, agile and free and open source software development. The authors noted that these three development approaches had a lot in common although homonyms in the naming of practices projected huge differences (Magdaleno et al., 2012). We complement these authors' research and explore those practices that have been carried over from the Spiral model (Boehm, 1986) into Scrum and Extreme Programming (XP).

In an effort to collate agile practices used to build quality into the designed software, Arcos-medina and Mauricio (2019) conducted a systematic literature review to understand how quality was handled in agile methods. They identified the practices, how the practices enabled quality in the developed software and how the quality impact was measured. Such a study would be related to ours in the sense that we seek to check whether the current agile methods incorporate quality practices that originated from the traditional methods. In the next section we discuss how we conducted the study.

3. Method

We carried out an analysis of three software development methodologies. We were interested in identifying quality practices that were designed after 1970, as this is the year that saw the advent of the historic Waterfall model (Boehm, 2006). The Spiral model was used as the source of these historic practices, since it was designed to incorporate practices from the Waterfall model and Prototyping (Boehm, 1986; Pressman & Maxim, 2015). A practice was deemed resilient if it was incorporated into the Spiral model and subsequently into any of the two representative agile methods, Scrum (Schwaber, 1997) and XP (Beck, 1999).

We created a table to capture the quality practices. First we captured those practices that were discussed in the Spiral model. We did not follow the methodologies stages in capturing the stages after noting that in some cases a practice appearing early in one method would appear later in another method, or worse still would appear repeatedly in a number of phases. A good example of such a practice is prototyping which appears during requirements elicitation in Scrum, and appears in XP as spike solutions during coding. Developers create spike solutions as per need to fully understand how code would behave in a given environment. We then analysed the two methods separately, capturing those practices that were projecting the same principle as a practice in the Spiral model. In the next section, we present the results of our analysis of the practices identified in three methods.

4. Results

Table 1 shows the practices that appear in the Spiral model, and still appear either in Scrum or XP or both.

*Table 1: Quality practices from participating methods*

| Traditional Methods Practices/Techniques | | Agile Methods Practices/Techniques | |
|---|---|---|---|
| **Spiral Model (Boehm, 1986)** | **Transition period before agile** | **Scrum (Schwaber,1997; Schwaber& Sutherland, 2013)** | **XP (Beck, 1999)** |
| i. Prototyping/simulation | | | Spike solution |
| ii. Incremental development | | Incremental development | Incremental development |
| iii. End of cycle review | | Sprint review | |
| iv. Successor phase planning | Practice refinement or elimination | Sprint Planning | Release planning |
| v. Iterative (go back to earlier stages) | | Iterative development | Iterative development, customer driven |
| vi. Risk-driven development, defer low risk modules | | Continuous risk assessment | Risk driven development |
| vii. Software automation | | | |
| viii. Software reuse | | | |
| ix. Minimize resource usage | | | |
| x. Critical design review | | | |
| xi. Risk planning | | Risk planning | |

Practices that have been carried over to agile methods include: prototyping, incremental development, successor stage/phase planning, end of cycle review, iterative development, risk driven development and risk planning. Prototyping is an old practice in software engineering. The practice first appeared in the traditional methods of software development as a means of understanding user requirements (Batool et al., 2013; IEEE, 2006; Strode & Clark, 2007). In the Waterfall model its main role was to understand requirements in the form of throw away prototyping (Strode & Clark, 2007). In code driven development methods like XP, the practice has assumed a new name, that of spike solution. However, the principle is still the same; to fully understand the problem, and test whether the suggested solution (code) for the problem works.

Another key practice that emerged during the traditional approach to software development and has earned a place in the agile methods is incremental development. Incremental development is closely linked to iterative development and modular design. Incremental development is a practice that ensures that the changing environment through the development process is taken care of (Beck, 1999; Boehm, 1986; Schwaber, 1997; Schwaber & Sutherland, 2014). As the software is built in increments, this ensures transparency in the development process, ensuring quality software as each increment is tested for meeting user requirements and smooth integration with the already delivered increments (Haider et al., 2021). This means, the first deliverable will be tested $n$ times if there are $n$ increments.

Next phase planning is unavoidable in an iterative development environment. Since the software is delivered in increments, such planning enables the software team, to set quality targets for the increment, at the same time ascertaining that the team is still developing the kind of software that the user wants. In Scrum, this entails unpacking the backlog items and establishing their requirements (Schwaber & Sutherland, 2014). In XP this means working with the customer and keeping everything simple to ensure development velocity (Shrivastava et al., 2021).

End of cycle review as projected in Scrum enables stakeholders to assess progress made in the just concluded spiral, and to ascertain commitment for the next spiral (Boehm, 1986). In Scrum this is called sprint review. This is also a short stakeholders' meeting in Scrum to review progress in the just ended sprint, at the same time making any adaptations to the product backlog in case of any need (Schwaber & Sutherland, 2014).

Risk management and risk planning seems to be an important aspect in software development. The Spiral model is considered a risk-based development approach (Boehm, 1986; Pressman & Maxim, 2015). High-risk components are analysed in depth. A risk plan is created at the onset of the development process, and the high risk components of the software get worked on ahead of the low risk components. At every iteration, the risks are reviewed and the risk plan adjusted accordingly. In the other methods practices like pair programming, on-site customer and small-release plans are seen as risk mitigation measures.

5. Discussion

The results of our review of the three software development methodologies chosen to represent the two important phases in the progression of software development show that, there are some practices that are generic in the development of software, regardless of the philosophy of the methodology. A prototype helps designers to fully understand the problem, at the same time affording them the opportunity to test the feasibility of the proposed solution. While the debate among agilists and the traditionalists has been on heavy documentation on the side of traditionalists as opposed to light documentation for the agilist; and too much rigidity in the processes for the traditional developer as opposed to the flexible agile developer, it is clear that the old standing software quality practices such as prototyping, can be incorporated into either side for the delivery of quality software.

Prototyping has also been incorporated into other agile methods not considered in this study. It is a practice in methods like Dynamic software development method (DSDM), Adaptive software development method (ASD) (Abrahamsson et al., 2002), and Secure-solo software development methodology Secure –SSDM (Moyo & Mnkandla, 2020) among others. Schön et al. (2017) noted that prototyping was one of the methods used to enable a quality requirements engineering process. It is clear in this case that prototyping is a practice that has stood the test of time, and can be carried out at any stage of the development process.

Similarly developing software in increments has proven to be key in the success of software development. Whether the end result of the incremental cycle is termed a component (Boehm, 1986) or a backlog (Schwaber & Sutherland, 2014), or user story (Beck, 1999), what is key is that developers breakdown the project into small pieces for manageable processes. The principle is still the same, minimise risk by developing the project in small increments that the stakeholders can assess at the end of each increment.

While some practices such as critical design review seem to have disappeared in the transition period, this may not necessarily be the case. Concepts of such practices have been engulfed in other practices which address a similar concept. Similarly, software reuse, may not necessarily be projected in the methodologies reviewed, but it is an inherent practice of most object-oriented programming languages.

6. Conclusion

In this paper our aim was to identify software development practices that promote quality in the developed software. Our aim was to show that while method engineers adopt various philosophies in the design of their methods, there is always a meeting point in their methods. That of the quality practices meant to build quality into their software products. Such practices remain the cornerstone of software development and are worth considering when designing software development methods or developing software in an ad hoc manner.

Our study considered three methods, further studies could consider more methods for the purposes of creating a pool of the long-standing quality practices that could serve as a source for future method engineers and software developers.

**References**

Abrahamsson, P., Salo, O., Ronkainen, J., & Warsta, J. (2002). Agile software development methods: Review and analysis. In *Espoo 2002, VTT Publications*. https://doi.org/10.1076/csed.12.3.167.8613

Arcos-medina, G., & Mauricio, D. (2019). Aspects of software quality applied to the process of agile software development : a systematic literature review. *International Journal of System Assurance Engineering and Management*, *10*(5), 867–897. https://doi.org/10.1007/s13198-019-00840-7

Batool, a, Motla, Y. H., Hamid, B., Asghar, S., Riaz, M., Mukhtar, M., & Ahmed, M. (2013). Comparative study of traditional requirement engineering and Agile requirement engineering. *Advanced Communication Technology (ICACT), 2013 15th International Conference On*, 1006–1014.

Beck, K. (1999). Embracing change with extreme programming. *Computer*, *32*(10), 70–77.

Boehm, B. (1986). A Spiral Model of Softwre Development and Enhancement. *ACM SIGSOFT Software Engineering Notes*, *11*(4), 21-.

Boehm, B. (2006). A view of 20th and 21st century software engineering. *Proceedings of the 28th International Conference on Software Engineering SE - ICSE '06*, 12–29. https://doi.org/doi: 10.1145/1134285.1134288

García-Mireles, G. A., Moraga, M. Á., García, F., & Piattini, M. (2015). Approaches to promote product quality within software process improvement initiatives: A mapping study. *Journal of Systems and Software*, *103*, 150–166.

Haider, M., Riesch, M., & Jirauschek, C. (2021). Realization of best practices in software engineering and scientific writing through ready-to-use project skeletons. *Optical and Quantum Electronics*, *53*(10), 1–17. https://doi.org/10.1007/s11082-021-03192-4

IEEE. (2006). *IEEE Standard for Developing a Software Project Life Cycle Process* (Vol. 2006, Issue July). https://doi.org/10.1109/IEEESTD.2006.219190

Karhapää, P., Behutiye, W., Rodríguez, P., Oivo, M., Costal, D., Franch, X., Aaramaa, S., Choraś, M., Partanen, J., & Abherve, A. (2021). Strategies to manage quality requirements in agile software development: a multiple case study. *Empirical Software Engineering*, *26*(2). https://doi.org/10.1007/s10664-020-09903-x

Kassab, M. (2015). The Changing Landscape of Requirements Engineering Practices over the Past Decade. *IEEE International Workshop on Empirical Requirements Engi- Neering (EmpiRE)*, *January*, 1–8. https://doi.org/10.1109/EmpiRE.2015.7431299

Magdaleno, A. M., Werner, C. M. L., & Araujo, R. M. De. (2012). Reconciling software development models: A quasi-systematic review. *Journal of Systems and Software*, *85*(2), 351–369.

Maxim, B. R., & Kessentini, M. (2016). An introduction to modern software quality assurance. In *Software Quality Assurance* (pp. 19–46). Elsevier Inc. http://dx.doi.org/10.1016/B978-0-12-802301-3.00002-8

Mbambo-Thatha, B., & Moyo, S. (2014). Towards non-gendered ICT education: The hidden curriculum at the National University of Science and Technology in Zimbabwe. In I. Buskens & A. Webb (Eds.), *Women and ICT in Africa and the Middle East: Changing Selves, Changing Societies* (pp. 45–55). Zed Books Ltd.

Moyo, S. (2020). *A Software Development Methodology for Solo Software Developers: Leveraging the Product Quality of Independent Developers* (Issue February). University of South Africa.

Moyo, S., & Mnkandla, E. (2020). A Novel Lightweight Solo Software Development Methodology with Optimum Security Practices. *IEEE Access*, *8*, 33735–33747. https://doi.org/10.1109/ACCESS.2020.297100

Pressman, R., & Maxim, B. (2015). *Software Engineering: A Practitioner's Approach* (8th

ed.). McGraw Hill,Education.

Schön, E. M., Thomaschewski, J., & Escalona, M. J. (2017). Agile Requirements Engineering: A systematic literature review. *Computer Standards and Interfaces*, *49*, 79–91. https://doi.org/10.1016/j.csi.2016.08.011

Schwaber, K. (1997). SCRUM Development Process. In *Business Object Design and Implementation* (pp. 117–134). Springer. http://link.springer.com/10.1007/978-1-4471-0947-1_11

Schwaber, K., & Sutherland, J. (2014). The Scrum Guide. In *Scrum.Org and ScrumInc*. https://doi.org/10.1053/j.jrn.2009.08.012

Shrivastava, A., Jaggi, I., Katoch, N., & Gupta, D. (2021). A Systematic Review on Extreme Programming. *Journal of Physics*, *1969*(2021), 1–11. https://doi.org/10.1088/1742-6596/1969/1/012046

Strode, D., & Clark, J. (2007). Methodology in software development capstone projects. *20th Annual Conference of the NACCQ, Nelson, NZ*, 243–251. http://www.citrenz.ac.nz/conferences/2007/243.pdf